\documentstyle[11pt,paspconf,epsf]{article}

\begin{document}

\title{Are Gamma-Ray Bursts Signals of Supermassive Black Hole Formation?}
\author{K. Abazajian, G. Fuller, X. Shi}
\affil{Department of Physics, University of California, San Diego,
       La Jolla, California 92093-0350}

\begin{abstract} 

The formation of supermassive black holes through the gravitational
collapse of supermassive objects ($\rm M \ga 5 \times 10^4\,\rm
M_\odot$) has been proposed as a source of cosmological $\gamma$-ray
bursts.  The major advantage of this model is that such collapses are
far more energetic than stellar-remnant mergers.  The major drawback
of this idea is the severe baryon loading problem in one-dimensional
models.  We can show that the observed ${\rm log}\, N - {\rm log}\, P$
(number vs. peak flux) distribution for gamma-ray bursts in the BATSE
database is not inconsistent with an identification of supermassive
object collapse as the origin of the gamma-ray bursts.  This
conclusion is valid for a range of plausible cosmological and
$\gamma$-ray burst spectral parameters.

\end{abstract}

\section{Introduction}

We investigate aspects of a recent model for the internal engine powering
$\gamma$-ray bursts (GRBs).  This model produces a GRB fireball
through neutrino emission and annihilation during the collapse of a
supermassive object into a black hole (Fuller \& Shi, 1998).  This
supermassive object may either be a relativistic cluster of stars or a
single supermassive star.  The collapse of a supermassive object
provides an exceedingly large amount of energy to power the burst,
much higher than the amount of energy available in stellar remnant
models. In addition, the rate of collapse of these objects--when
associated with galaxy-type structures--is similar to the GRB rate.

The recent observations of afterglows and galaxies associated with
GRBs have secured that at least some have a cosmological origin and
therefore must be extremely energetic events. The inferred redshifts
of GRB 970508 and GRB 971214 suggest isotropic emission energies of
${\sim}10^{52}$ and $3 \times 10^{53}$ ergs (Metzger et al. 1997a,
1997b; Kulkarni et al.  1998). The energy in $\gamma$-rays alone for
GRB 971214 is equivalent to $16\%$ of the rest mass of the
sun. Producing this amount of energy in $\gamma$-rays is difficult for
stellar remnant collapse models, where the total amount of
gravitational binding energy released in a ${\sim}1\, M_{\odot}$
configuration is ${\sim}10^{54}$ ergs (Wijers et al.  1998).  However,
{\it if} a stellar-remnant collapse manages to concentrate energy
deposition into ${\sim}1\%$ of the sky, then it {\it may} produce a
burst with the observed energies.

Supermassive black holes are abundant in the universe.  They are
inferred to power active galactic nuclei (AGNs) and quasars; every
galaxy examined so far seems to possess a supermassive black hole in
its center (van der Marel et al. 1997). These black holes could have
had supermassive objects as progenitors (Begelman \& Rees 1978). Two
venues of supermassive black hole formation are considered: in one, a
dense cluster of $1\, M_\odot$ stars is disrupted by collisions and
coalesces into a central supermassive star; in the second venue, the
cluster as a whole undergoes a post newtonian collapse into a
supermassive object. A supermassive star may also be formed directly
through the collapse of a ${\sim}10^5\, M_\odot - 10^6\, M_\odot$
primordial gas cloud when cooling in these is not efficient (Peebles
\& Dicke 1968 and Tegmark et al. 1997).

We will discuss the formation of a GRB fireball in both collapse
scenarios. We will also address how the highly variable time structure
of GRBs can occur in supermassive object collapse, and how the
fireball can avoid ``baryon-loading'', which can incapacitate the
formation of a relativistic fireball. It should be noted that
Prilutski \& Usov (1975) previously described the emission of a GRB
from magneto-energy transfer during collapse of supermassive rotators
$({\sim}10^6\, M_\odot)$ believed to power AGNs and quasars.  The
neutrino energy transfer process we discuss is not necessarily tied to
AGNs or quasars, but to the formation of the black holes powering
them.

\section{Fireballs from Supermassive Black Hole Formation}

\subsection{Supermassive Star Collapse}
In the first venue of supermassive black hole formation, a
supermassive star undergoes a general relativistic
(Feynman-Chandrasekhar) instability.  A core of mass $M_5^{\rm
HC}\equiv M^{\rm HC}/10^5M_\odot$ collapses homologously and drops
through the event horizon, releasing a gravitational binding energy of
$\sim E_{\rm s}\approx 10^{59}M^{\rm HC}_5\,{\rm erg}$.  The mass of
the homologous core can be an order of magnitude (or more) less than
the mass of the initial hydrostatic supermassive star, $M_5^{\rm
init}\equiv M^{\rm init}/10^5M_\odot$.

During collapse, neutrinos are thermally emitted due to $e^\pm$
annihilation in the core.  The luminosity of the neutrinos goes as the
ninth power of the core temperature (Dicus 1972).  This luminosity can
be approximated from the product of neutrino emissivity (Schinder et
al. 1987; Itoh et al. 1989) near the black hole formation point and
the volume inside the Schwarzschild radius, $4\times
10^{15}\,(T_9^{\rm Schw})^9\,(4\pi r_{\rm s}^3/3)$ erg/sec, where
$T_9^{\rm Schw}$ is the characteristic average core temperature in
units of $10^9$ K at the black hole formation epoch. In a spherical
nonrotating supermassive star this is
\begin{equation}
T_9^{\rm Schw} \approx 12 {\alpha}_{\rm Schw}^{1/3}
{\left({{11/2}\over{g_{\rm s}}}\right)}^{1/3} {\left( {{M_5^{\rm
init}}\over{M_5^{\rm HC}}}\right)}^{1/6} {\left( M_5^{\rm HC}
\right)}^{-1/2}.
\label{temp}
\end{equation}
Here ${\alpha}_{\rm Schw}$ is the ratio of the final entropy
per baryon to the value of this quantity in the initial
pre-collapse hydrostatic star,
and $g_{\rm s} \approx g_b + 7/8 g_f \approx 11/2$ is the
statistical weight of relativistic particles in the core.  The
characteristic free-fall timescale is labelled $t_{\rm s}\approx
M^{\rm HC}_5\,{\rm sec}$,
and the characteristic radius (the Schwarzschild radius)
is $r_{\rm s}\approx 3\times 10^{10}M^{\rm HC}_5\,{\rm cm}$. It has
been shown that the ratio of the homologous core mass to the initial
mass is $M_5^{\rm HC}/M_5^{\rm init}\approx \sqrt{2/5.5}{\alpha}_{\rm  
Schw}^{2}$ (Fuller, Woosley \& Weaver 1986), so that $T_9^{\rm
Schw}\approx 13(M^{\rm HC}_5)^{-1/2}$.
The neutrino luminosity is
\begin{equation}
L_{\nu\bar\nu}\sim 4\times 10^{15}\,(T_9^{\rm Schw})^9\,(4\pi\,
r_{\rm s}^3/3)\,{\rm erg/sec}
\approx 5\times 10^{57} (M^{\rm HC}_5)^{-3/2}\,{\rm erg/sec}.
\label{luminosity}
\end{equation}
About 70\% of the neutrino emission will be in the $\nu_e\bar\nu_e$
channel (Woosley, Wilson \& Mayle 1986). 

This ample $\nu\bar\nu$ emission can create a fireball above the core
through $\nu\bar\nu\rightarrow e^+e^-$.  The neutrino luminosities
will undergo gravitational redshift, which depresses energy deposition
above the star; however, this will be compensated by increased
$\nu\bar\nu$-annihilation from gravitational bending of null
trajectories (Cardall \& Fuller 1997). The neutrino emission is nearly
thermal (Shi \& Fuller 1998), allowing the neutrino energy deposition
rate to be approximated as
\begin{equation}
\dot{Q}_{\nu\bar\nu}(r)\sim 4\times 10^{22}\,(M^{\rm HC}_5)^{-7.5}
(r_{\rm s}/r)^8\,{\rm erg}\,{\rm cm}^{-3}{\rm s}^{-1}.
\end{equation}
The total energy injected into the fireball above a radius $r$ by this
process is
\begin{equation}
E_{\rm f.b.}(r)=t_{\rm s}\,
{\int_r^\infty} 4\pi r^2\dot{Q}_{\nu\bar\nu}(r){\rm d}r
\sim 2.5\times 10^{54}\,(M^{\rm HC}_5)^{-3.5}(r_{\rm s}/r)^5\,
{\rm erg}.
\label{fireball}
\end{equation}
This is an unequivocally large amount of energy. For a star
where $M_5^{\rm HC}=0.5$, the energy of the fireball will be
$\sim10^{53}$ erg at a radius $r\sim 3r_{\rm s} \sim 10^{11}\,{\rm
cm}$.  This would correspond to the energy of a GRB with isotropic
emission at a redshift $z\approx 3$.

\subsection{Supermassive Star Cluster Collapse}

The second venue for the production of a GRB fireball during
supermassive black hole formation is the collapse of a star cluster of
$10^5 - 10^9\, M_\odot$. The cluster undergoes a general relativistic
instability (Shapiro \& Teukolsky 1985) where collisions of $M_* \sim
M_\odot$ stars could produce the neutrino emission powering a
fireball. During the collapse, the stars will have relativistic speeds
($\Gamma \sim 1$) and a zero impact parameter collision of a pair will
produce a typical entropy per baryon of $S\sim
10^4\Gamma^{1/2}(g_s/5.5)^{1/4}
(M_\odot/M_*)^{1/4}(V_*/V_\odot)^{1/4}$ with $T_9\sim 1$, and where
$V_*/V_\odot$ is the ratio of the stellar collision interaction volume
to the solar volume.  Generally, the collisions will have a non-zero
impact parameter, and involve the less dense outer layers of the star,
where there will be larger entropies. These entropies could be high
enough ($S\sim 10^7$) to produce the pair fireball without the need
for neutrino heating.  The complex structure and baryon-free regions
between stars can provide areas for fireballs to form with low
baryon-loading.  Both the collisions and neutrino emission are
stochastic processes which might lead to the complex time structure of
GRBs.

\section{Event Rates and the ${\rm log}\, N - {\rm log}\, P$ Distribution}

If all collapses occur at a single redshift, $z$, the observed rate is
\begin{equation}
4\pi r^2a_z^3{{\rm d}r\over {\rm d}t_0}{\rho_{\rm b}F(1+z)^3
\over M^{\rm init}},
\end{equation}
where $r$ is the Friedman-Robertson-Walker comoving coordinate
distance of the objects, $a_z$ is the scale factor of the universe
corresponding to $z$ (with $a_0=1$), $t_0$ is the age of the universe,
$\rho_{\rm b}\approx 2\times 10^{-29}\,\Omega_{\rm b}h^2\,{\rm
g}\,{\rm cm}^{-3} \approx 5\times 10^{-31}{\rm g}\,{\rm cm}^{-3}$
(Tytler \& Burles 1997) is the baryon density of the universe, $h$ is
the Hubble parameter in 100 km$\,{\rm s}^{-1}\,{\rm Mpc}^{-1}$, and
$F$ is the fraction of all baryons in supermassive objects.  With
collapses occuring at $z\sim 3$ we have $r\sim 3000h^{-1}$ Mpc, and
the corresponding collapse rate is
\begin{equation}
0.15F\,(M_5^{\rm init})^{-1}\,{\rm sec}^{-1}
\sim 10^4F\,(M_5^{\rm init})^{-1}\,{\rm day}^{-1}.
\end{equation}
With $F\sim 0.1\%$, and with a 100\% detection efficiency, we will observe
one collapse per day, assuming isotropic emission. This corresponds to a
density of supermassive black holes of $7\,h^2/{\rm Mpc^3}$, about
$350\, h^{-1}$ per $L_*$ galaxy, or $\la 10\, h^{-1}$ per galaxy-scale
object (i.e. including dwarf galaxies).  

It is instructive, however, to estimate the rate of supermassive
object collapse in terms of numbers of Lyman limit systems and damped
Ly$\alpha$ systems. Employing a column denisty $N_{\rm HI}$
distribution per unit column density per unit absorption distance of
$10^{13.9} N^{-1.74}_{\rm HI}$ (Storrie-Lombardi, Irwin \& McMahon
1996), the rate of supermassive object collapse will be comparable to
that of GRBs if every Ly$\alpha$ system with $N_{\rm HI} \ga 10^{18}\,
{\rm cm}^{-2}$ experiences one supermassive object collapse.

The relation of the number of GRBs with peak flux (${\rm log}\, N-{\rm
log}\, P$) may be able to tell us something about the distribution of
GRB sources in the universe.  To see if there may be a correlation
between this distribution and that of supermassive object collapse, we
count the number of supermassive objects through the quasar
population. That is, if quasars have some characteristic lifetime,
then we can say that the comoving supermassive object number density
is proportional to the comoving quasar number density. There has been
some recent work in the evolution of the number density of quasars
(Maloney \& Petrosian 1998, Shaver et al. 1998).  We use the evolution
of quasar number to sum the number of supermassive collapse events for
a standard candle and various cosmologies.  The peak flux distribution
calculated from supermassive object collapse is not inconsistent with
the GRB ${\rm log}\, N-{\rm log}\, P$ relation, considering the
uncertainties in the quasar epoch and GRB luminosity distribution.  In
Figure 1, we show the case for $\Omega_m = 0.25, \Omega_\Lambda = 0,
\alpha = 0.7, z_{\rm th} = 3.4$, where $\alpha$ is the GRB spectral
index, and $z_{th}$ is the cutoff redshift of the BATSE detector. We
must be careful, however, since the luminosity function of GRBs is
unknown, and recent work has shown that this can affect the observed
peak flux distribution greatly (Krumholz et al. 1998).

\begin{figure}
\plotone{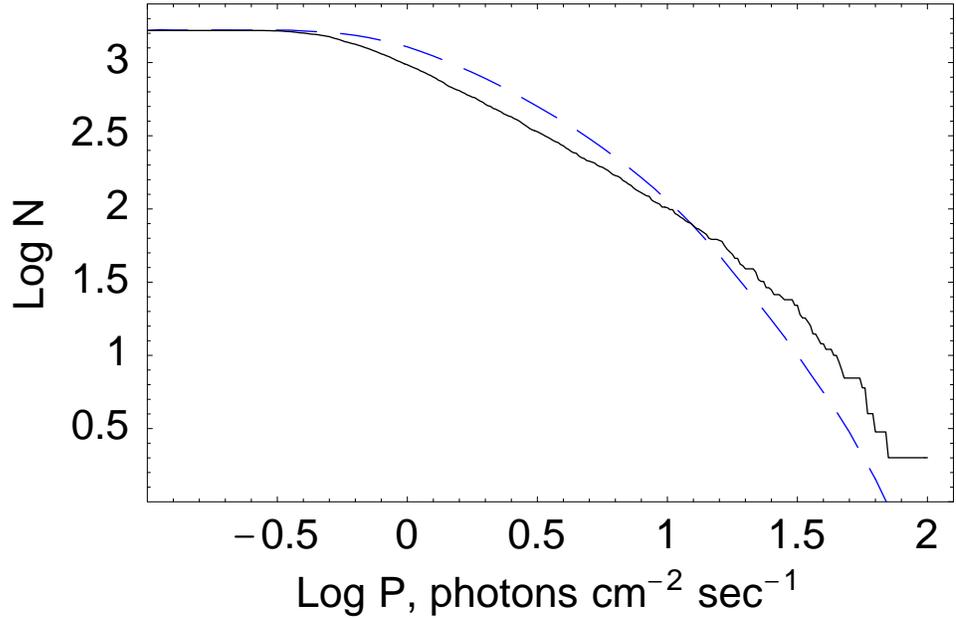}
\caption{The BATSE 4B Catalogue number versus peak flux (solid) and
the expected distribution (dashed) for supermassive object collapse
associated with quasars.  Here $\Omega_m = 0.25, \Omega_\Lambda = 0,
\alpha = 0.7, z_{\rm th} = 3.4$.}
\end{figure}

\section{GRB Time Structure and Baryon Loading}

The quickly varying time structure of GRBs limits the size of the
region powering the fireball to the distance light can travel during
this time variation, ${\sim}10^7$ cm (Piran 1998). In the first venue
described above of GRB production (supermassive star collapse) the
characteristic size of the emission region in this spherically
symmetric model is the Schwarschild radius, ${\sim}10^{10}$ cm.  The
supermassive star's collapse and the formation of fireball(s) will not
generally be spherically symmetric, nor will convective processes be
unimportant.  So, we can say that the above energy scales will
be deposited and localized by neutrino annihilation into regions
outside of the core that are a fraction of the core's size.  These
regions will be distributed around the core, and may produce
variability through superpositions and instabilities.  In the second
venue, $1\, M_\odot$ stars have the same physical scale,
${\sim}10^{10}$ cm, but can produce variability in the means described
in \S 2.3 or through a localization of neutrino annihilation.

Producing a small region of high photon energy density and high
entropy per baryon---the GRB fireball---is a challenge for all GRB
models.  The supermassive star model, in the one-dimensional case,
also does not deposit the needed energies in a ``baryon-free'' region.
However, if some of the tremendous energy deposited by the collapse
does find itself in an area with low baryon density, it will produce a
GRB fireball of large energies.  This can happen when the star is
rotationally flattened, where the neutrinos deposit their energy along
the ``baryon-free'' axis of rotation.

\acknowledgments

I would like to thank my parents, Nazar and Raisa, whose support made
my attendance at this symposium possible.  I would also like to thank
the hospitality of Hasmig Haroutoonian and the Local Organizing
Committee.  This research is supported by NASA grant NAG5-3062 and NSF
grant PHY95-03384 at UCSD.

\end{document}